\documentclass[aps,superscriptaddress,twocolumn,showpacs]{revtex4-1}
\pdfoutput=1
\usepackage{color}
\usepackage{amsmath, amsthm, amsfonts}    % need for subequations
\usepackage{amssymb}
\usepackage{mathtools}
\usepackage{mathptmx} 
\usepackage[table]{xcolor}
\usepackage{dsfont}
\usepackage{enumitem} 
\usepackage{appendix}
\usepackage{braket}
\usepackage{pifont}   % for better checkmark and cross

\usepackage[skins,theorems]{tcolorbox}
\tcbset{highlight math style={enhanced,
  colframe=red,colback=white,arc=0pt,boxrule=1pt}}
\usepackage{cancel}
\usepackage{empheq}
\usepackage{lipsum}% http://ctan.org/pkg/lipsum
\usepackage{graphicx}% http://ctan.org/pkg/graphicx
\usepackage{tikz}
\usepackage{makecell}
\usepackage[unicode=true,pdfusetitle, bookmarks=true,bookmarksnumbered=false,bookmarksopen=false,
 breaklinks=true,pdfborder={0 0 0},backref=false,colorlinks=true,citecolor=blue]{hyperref}
\usepackage{graphicx}   % need for are exact and remain valid regardless of multi- pole ORes
\usepackage[caption=false]{subfig}       % use for side-by-side are exact and remain valid regardless of multi- pole ORes and   to have captions justified and small work size
\usepackage{bm}            % bold math
\usepackage[normalem]{ulem}  % use for underlining

\newlength\imagewidth
\newlength\imagescale
\def\be{\begin{eqnarray}}
\def\ee{\end{eqnarray}}

\newcommand{\til}[1]{{\overline{#1}}}

\def\E{{\bf E}}
\def\H{{\bf H}}
\def\B{{\bf B}}

\def\p{{\bf p}}
\def\m{{\bf m}}

\usepackage{calrsfs}
\DeclareMathAlphabet{\mathcal}{OMS}{cmsy}{m}{n}
\def\im{{\rm i}}

\definecolor{JOT-color}{named}{blue}
\definecolor{CSF-color}{named}{orange}
\newtcolorbox{keyresultbox}{
  colback=white!10, % background color
  colframe=black,  % bORer color
  boxrule=0.5pt,
  arc=2pt,         % rounded corners
  left=6pt, right=6pt, top=6pt, bottom=6pt
}
\begin{document}

\title{An Exact Energy Conservation Law for Magneto-Optical Nanoparticles}

%\title{Energy Conservation Dictates Forbidden and Strong Magneto-Optical Dipolar Scattering}

%\title{An Exact Energy Conservation Law for Magneto-Optical Nanoparticles}

\author{Jorge Olmos-Trigo}
\email{jolmostrigo@gmail.com}
\affiliation{Departamento de Física de Materiales, Universidad Autónoma de Madrid, 28049
Madrid, Spain.}

\begin{abstract}
Energy conservation imposes fundamental bounds on the polarizabilities of nanoparticles (NPs). While such bounds are well established for isotropic and bianisotropic NPs, they remain unexplored for magneto-optical NPs. Here, we derive the exact energy-conservation law governing the electric and magnetic dipolar response of axially symmetric magneto-optical NPs under general illumination conditions and arbitrary external magnetic fields. Two central results follow from energy conservation: (i) purely magneto-optical scattering, where the non-magnetic polarizability vanishes, is fundamentally forbidden, and (ii) strong magneto-optical scattering regimes, in which the magneto-optical polarizability dominates, are intriguingly allowed. %Unfortunately, the standard approximation of truncating the polarizability tensor to linear order in the external magnetic field masks these results.

\end{abstract}

\maketitle

{\emph{Introduction.}}---Conservation of energy is one of the most fundamental laws in physics~\cite{noether1918invariante}. In nanophotonics, this law imposes strict bounds on the polarizabilities of nanoparticles (NPs). 
While these bounds have been thoroughly studied for isotropic~\cite{hulst1957light,rahimzadegan2020minimalist,olmos2020optimal,olmos2020unveiling,olmos2024revealing,olmos2024solving} and chiral~\cite{jaggard1979electromagnetic,sersic2011magnetoelectric,albooyeh2016purely} NPs, the effect of an external magnetic field on such constraints has received comparatively little attention~\cite{Albaladejo2010}. 
This gap is particularly striking for one of the most commonly employed magneto-optical systems: axially symmetric magneto-optical NPs~\cite{ott2018circular, berger2015enhanced, arruda2016electromagnetic, kuttruff2021magneto, safaei2024optical, kort2013tuning, cheng2020light,  christofi2018metal, ying2018strong, gabbani2022high, gonzalez2024enhanced}.  
For such NPs, the electromagnetic response  is fully described by four scalar polarizabilities: the ordinary electric ($\rm{E}$) and magnetic ($\rm{M}$) polarizabilities, $\alpha_{\mathrm{OR,E}}$ and $\alpha_{\mathrm{OR,M}}$, and the magneto-optically induced polarizabilities, $\alpha_{\mathrm{MO,E}}$ and $\alpha_{\mathrm{MO,M}}$, which arise solely due to the externally applied magnetic field.
Yet, even in this scenario, an exact energy-conservation law remains elusive.

In this work, we derive such a conservation law, which sets strict constraints on the polarizabilities of magneto-optical NPs. Notably, these energy-conservation bounds hold for NPs made of both lossless and lossy materials, and must be met in any modeling of magneto-optical NPs. 
Furthermore, energy conservation reveals two fundamental results: i) if $\alpha_{\mathrm{OR},\beta}=0$ (for both $\beta = \rm{E}, \rm{M}$), energy conservation strictly enforces $\alpha_{\mathrm{MO},\beta}=0$, meaning that purely magneto-optical dipolar scattering cannot exist at the single-particle level; and ii) there are energy-preserving regimes in which $\alpha_{\mathrm{MO},\beta}$ significantly exceed $\alpha_{\mathrm{OR},\beta}$ even for relatively low values of the magnetic field. 
Importantly, truncating the polarizability tensor to linear order in the external magnetic field, an ubiquitous approximation in magneto-optics~\cite{lodewijks2014magnetoplasmonic,pakdel2012faraday,maccaferri2013tuning,pineider2013circular,marinchio2014magneto,vincent2011magneto,edelstein2019magneto}, obscures these two results.

Next, we introduce the setting used in this work~\footnote{Throughout, we adopt the following convention: a given quantity \(A\) may appear as \(A\), \(\bm{A}\), \(\til{A}\), or \(\tilde{A}\), corresponding respectively to its scalar, vector, 3x3 tensor, and 6x6 tensor forms}.

\begin{figure}[t!]
    \centering
\includegraphics[width=1\linewidth]{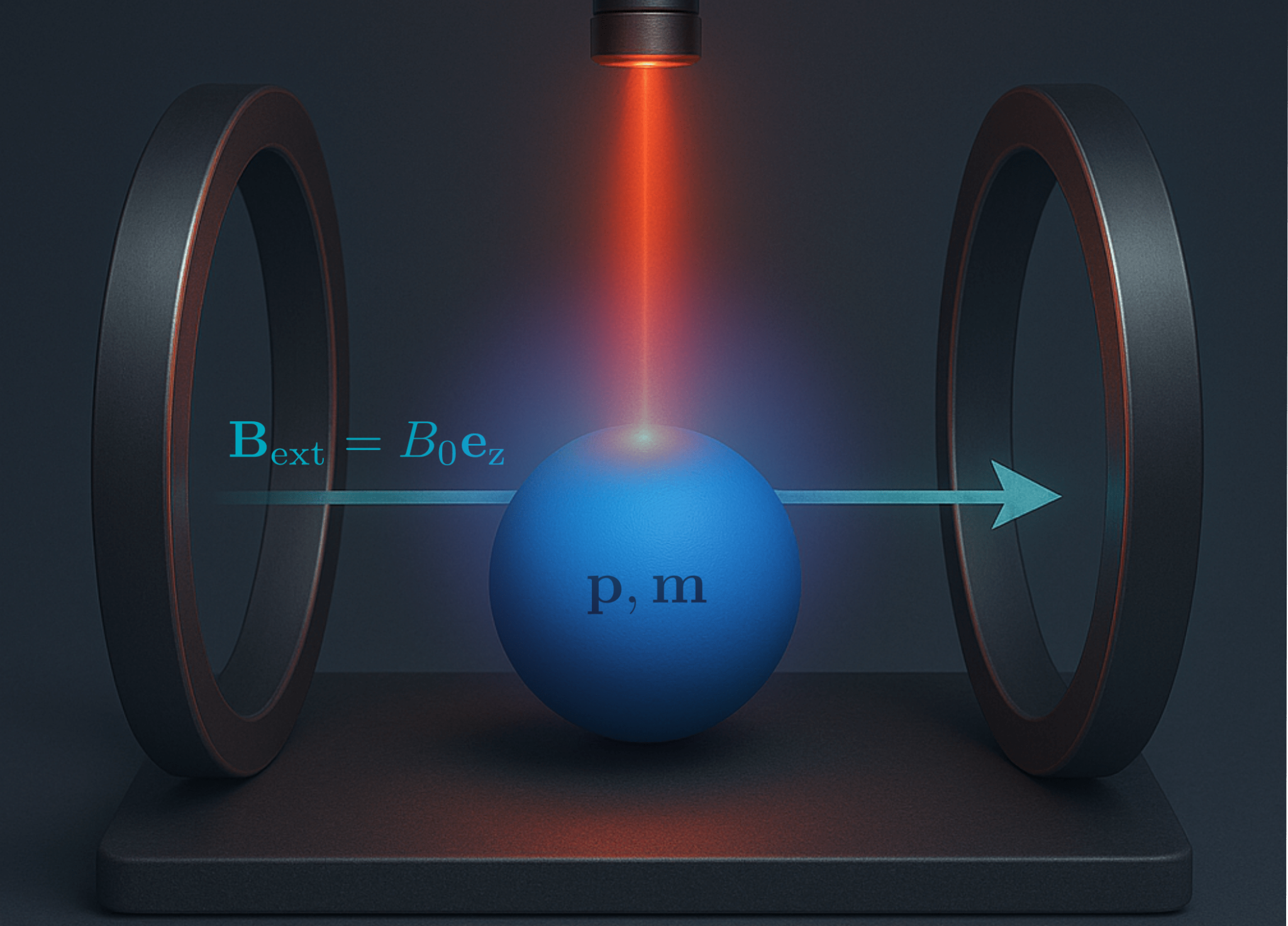}
\caption{A magneto-optical NP sustaining electric ($\p$) and magnetic ($\m$) dipolar modes is subjected to an external magnetic field $\B_{\rm{ext}}$.}
    \label{fig:placeholder}
\end{figure}

{\emph{The polarizability tensor of magneto-optical  NPs.}}--- We consider an axially-symmetric NP made of a homogeneous material with scalar permittivity $\varepsilon \in \mathbb{C}$ embedded in an otherwise homogeneous and isotropic medium with real permittivity $\varepsilon_{\rm{h}}$, refractive index $n_{\rm{h}} = \sqrt{\varepsilon_{\rm{h}}}$. In the presence of a static external magnetic field $\B_{\text{ext}}$, the scalar permittivity of the NP becomes a $3 \times 3$ tensor whose  components satisfy $\varepsilon_{ij}(\B_{\text{ext}}) = \varepsilon_{ji}(-\B_{\text{ext}})$. 
Without loss of generality, we assume  $\B_{\rm{ext}} = \pm B_0 \hat{\mathbf{e}}_{\rm{z}}$, as shown in Fig.~\ref{fig:placeholder}.
In addition to \(\mathbf{B}_{\rm{ext}}\), we consider an incident electromagnetic field \(\{\mathbf{E}, \mathbf{H}\}\) impinging on the magneto-optical NP.
This incident field induces an electric dipole 
\(\mathbf{p} = \varepsilon_{\rm{h}} \varepsilon_0 \til{\alpha}_{\rm{E}} \mathbf{E}\) 
and a magnetic dipole 
\(\mathbf{m} = \til{\alpha}_{\rm{M}}  \mathbf{H}\), 
where \(\til{\alpha}_{\rm{E}}\) and \(\til{\alpha}_{\rm{M}}\) are the electric and magnetic polarizability tensors in the presence of \(\mathbf{B}_{\rm{ext}}\). 
As is often done in the literature~\cite{scott2020enhanced}, we can embody all the previous information into a $6 \times 6$ polarizability tensor ${\tilde{\alpha}}$, namely,
\begin{align}\label{Rel}
\begin{pmatrix}
\p / \epsilon_h \epsilon_0 \\
 \m 
\end{pmatrix}
= \tilde{\alpha} \begin{pmatrix}
\E\\
\H
\end{pmatrix} = 
\begin{pmatrix}
\til{\alpha}_{\rm{E}} & \til{0}_{3x3}  \\
\til{0}_{3x3} & \til{\alpha}_{\rm{M}}  
\end{pmatrix}
\begin{pmatrix}
\E\\
 \H
\end{pmatrix},
\end{align}
where 
\be \label{pol_mag}
\til{\alpha}_\beta =
\left(
\begin{array}{ccc}
\alpha_{\rm{OR},\beta} & \mp \im \alpha_{\rm{MO}, \beta} & 0 \\
\pm \im \alpha_{\rm{MO}, \beta} & \alpha_{\rm{OR},\beta} & 0 \\
0 & 0 & \alpha_{\rm{IN},\beta}
\end{array}
\right).
\ee
Some comments are in order: In sharp contrast to most previous works~\cite{, lodewijks2014magnetoplasmonic,pakdel2012faraday,maccaferri2013tuning,pineider2013circular,marinchio2014magneto,vincent2011magneto, edelstein2019magneto}, we will retain the magnetic terms in  ${\alpha}_{\rm{OR},\beta}$, ensuring the validity of our results for objects exhibiting strong magneto-optical responses. 
The magneto-optical polarizability $\alpha_{\rm{MO}, \beta}$, in turn,  depends linearly on $\B_{\rm{ext}}$ and arises only when $\B_{\rm{ext}}$ is applied. That is, if $\B_{\rm{ext}} = 0$, then $\alpha_{\rm{MO}, \beta} = 0$, and $\tilde{\alpha}$ reduces to that of an axially symmetric NP, diagonal in its principal axes. Moreover, as we apply $\B_{\rm{ext}}$ in the z-direction, the polarizability  $\alpha_{\rm{IN},\beta}$ remains invariant under the action of the external magnetic field. 

Accordingly, the polarizability tensor we consider in Eq.~\eqref{pol_mag} is the most generic one for an axial-symmetric magneto-optical NP under the action of $\B_{\rm{ext}} = \pm B_0 \hat{\mathbf{e}}_z$.

Next, we derive the constraints that $\alpha_{\rm{OR},\beta}$, $\alpha_{\rm{MO}, \beta}$, and $\alpha_{\rm{IN},\beta}$ need to satisfy to guarantee the conservation of energy.

\emph{The energy conservation for magneto-optical  NPs.---} The energy conservation law for dipolar  NPs was introduced by Belov \emph{et al.} in 2003~\cite[Eq.~(10)]{belov2003condition}. Such a law be compactly written in our notation as:
\begin{equation} \label{belov}
\frac{\tilde{\alpha} - \tilde{\alpha}^\dagger}{2 \im} \geq  \frac{k^3}{6 \pi}   \tilde{\alpha}^\dagger \tilde{\alpha},
\end{equation}
where “${}^\dagger$” is the hermitian conjugate. The left-hand side of Eq.~\eqref{belov} is proportional to extinction, while the right-hand side is proportional to scattering~\cite{bohren2008absorption}.
In the absence of absorption, the inequality “$\geq$” in Eq.~\eqref{belov} becomes an equality “$=$”. 
Now, Eq.~\eqref{belov} can be conveniently expressed in terms of 
a positive semidefinite (PSD) hermitian matrix $\mathcal{\tilde{A}} = \mathcal{\tilde{A}}^\dagger$ as:
\begin{align} \label{jorge}
\mathcal{\tilde{A}}  \geq 0, \quad \text{where} \quad \mathcal{\tilde{A}} =  \frac{\tilde{\alpha} - \tilde{\alpha}^\dagger}{2 \im} -  \frac{k^3}{6 \pi} \tilde{\alpha}^\dagger \tilde{\alpha}.
\end{align}
To elucidate the role of the external magnetic field $\B_{\rm{ext}}$ in the conservation of energy for magneto-optical NPs, we expand the right-hand side of Eq.~\eqref{jorge} using Eq.~\eqref{pol_mag}.
After a little bit of algebra (see Appendix~\ref{Appendix A}), we arrive at: 
\begin{equation} \label{def}
\mathcal{\tilde{A}} = 
\begin{pmatrix}
\mathcal{\til{A}}_{\rm{E}} & \til{0}_{3x3}  \\
\til{0}_{3x3} & \mathcal{\til{A}}_{\rm{M}}
\end{pmatrix}, \quad \text{where} \quad
\mathcal{\til{A}}_\beta =
\begin{pmatrix}
\mathcal{J}_{\beta} & \mp \im \mathcal{O}_{\beta} & 0 \\
\pm \im \mathcal{O}_{\beta} & \mathcal{J}_{\beta} & 0 \\
0 & 0 & \mathcal{T}_{\beta}
\end{pmatrix}.
\end{equation}
In Eq.~\eqref{def}, we have introduced the real-valued  quantities:
\be \label{H11}
\mathcal{J}_{\beta} &=&  
\Im(\alpha_{\rm{OR},\beta})
- \frac{k^3}{6\pi} \left( |\alpha_{\rm{OR},\beta}|^2 + |\alpha_{\rm {MO}, \beta}|^2 \right),
\\  \label{H12}
\mathcal{O}_{\beta} &=& 
\Im(\alpha_{\rm {MO}, \beta})
- \frac{k^3}{3\pi} \Re(\alpha^*_{\rm{OR},\beta} \alpha_{\rm {MO}, \beta}),
\\  \label{H33}
\mathcal{T}_{\beta} &=& 
\Im(\alpha_{\rm{IN}, \beta})
- \frac{k^3}{6\pi} |\alpha_{\rm{IN}, \beta}|^2.
\ee
Here, $\Im(z)$ and $\Re(z)$ denote the imaginary and real parts of any complex number $z$. A few comments are in order. As Eqs.~\eqref{H11}-\eqref{H33} show, the electric dipole $\p$ and magnetic dipole $\m$ of magneto-optical NPs do not couple in the conservation of energy. In other words, energy conservation, as expressed in Eq.~\eqref{belov}, applies independently to each \(3\times3\) dipolar block, just as in the case of isotropic NPs sustaining electric and magnetic dipolar modes~\cite{bohren2008absorption}. This behavior stands in sharp contrast to that of bi-anisotropic NPs, where the electric and magnetic polarizabilities intertwine to ensure energy conservation~\cite{jaggard1979electromagnetic}.

Having noted these important points, we turn our attention back to $\mathcal{\tilde{A}}$, introduced in Eq.~\eqref{def}. The convenience of defining this $6\times 6$ tensor becomes now evident: if $\mathcal{\til{A}}_{\rm{E}} = \mathcal{\til{A}}_{\rm{M}} = 0$, then the  magneto-optical NP is lossless (and the converse statement also holds). In this lossless regime, we have $\mathcal{J}_{\beta}  = \mathcal{O}_{\beta}  = \mathcal{T}_{\beta} = 0  $, and hence,  the  polarizabilities necessarily satisfy: 
\be \label{Abs_1}
\Im(\alpha_{\rm{OR},\beta})
&=& \frac{k^3}{6\pi} \left( |\alpha_{\rm{OR},\beta}|^2 + |\alpha_{\rm{MO},\beta}|^2 \right), 
\\
\Im(\alpha_{\rm{MO},\beta})
&=& \frac{k^3}{3\pi} \Re(\alpha^*_{\rm{OR},\beta} \alpha_{\rm{MO},\beta}),
\\  \label{Abs_3}
\Im(\alpha_{\rm{IN},\beta})
&=& \frac{k^3}{6\pi} |\alpha_{\rm{IN},\beta}|^2.
\ee
Equations~\eqref{Abs_1}–\eqref{Abs_3} constitute the first important result of this work. Any physically consistent polarizability model must satisfy Eqs.~\eqref{Abs_1}–\eqref{Abs_3} for lossless magneto-optical NPs.
As a sanity check, we note that in the isotropic setting, where no external magnetic field is applied (\(\mathbf{B}_{\rm ext}=0\)) and the magneto-optical polarizabilities vanish (\(\alpha_{\rm MO,\beta}=0\)), the well-known energy conservation relations for lossless isotropic dipoles are recovered~\cite{olmos2024revealing}: \(\Im(\alpha_{\rm e}) = (k^3/6\pi)\,|\alpha_{\rm e}|^2\) and \(\Im(\alpha_{\rm m}) = (k^3/6\pi)\,|\alpha_{\rm m}|^2\). Here \(\alpha_{\rm e}\) and \(\alpha_{\rm m}\) are the electric and magnetic polarizabilities for $\B_{\rm{ext}} = 0$, respectively~\cite{olmos2019asymmetry, olmos2019enhanced}.

\begin{figure*}[t!]
    \centering
\includegraphics[width=1\linewidth]{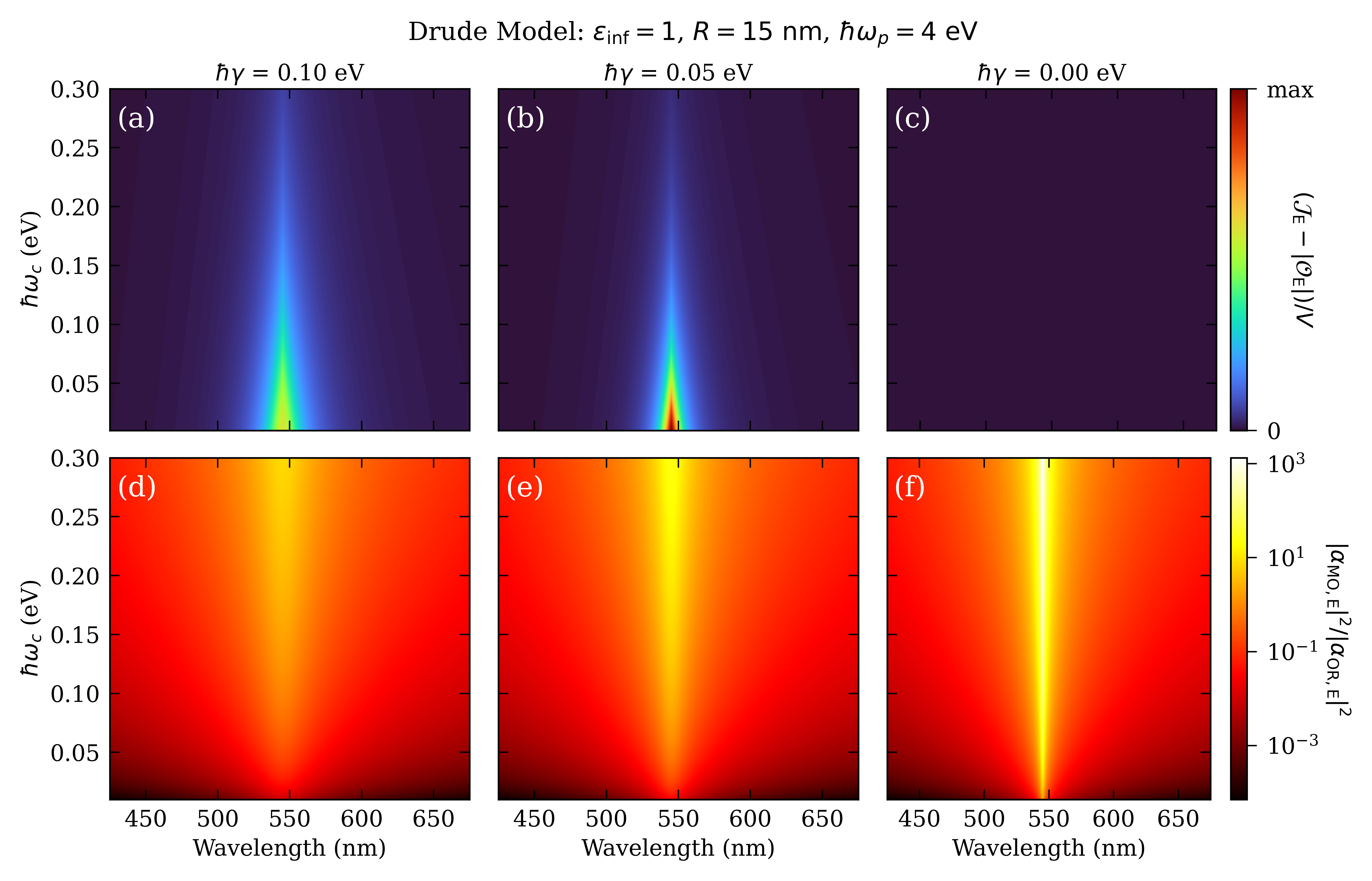}
   \caption{
Optical response of a spherical magneto-optical nanoparticle (radius $R = 15~\mathrm{nm}$) described by a Drude model with $\varepsilon_{\rm inf} = 1$ and $\hbar \omega_p = 4~\mathrm{eV}$. Top row: Conservation of energy given by $(\mathcal{J}_{\rm{E}} - |\mathcal{O}_{\rm{E}}|)/V$ for different damping rates $\hbar \gamma$. Here $V$ denotes the volume of the NP. Bottom row: logarithmic ratio $|\alpha_{\rm{MO}, \rm{E}}|^2 / |\alpha_{\rm{OR}, \rm{E}}|^2$. The incident wavelength $\lambda$ is indicated in nm, and $\hbar\omega_c$ in eV. Subplots are labeled (a)–(f).
}
    \label{fig}
\end{figure*}

In the following, we consider the general case imposed by  \(\mathcal{\tilde{A}} \geq 0\).

To determine the general constraints that $\mathcal{\tilde{A}} \geq 0$ imposes on the polarizabilities 
\(\alpha_{\rm OR,\beta}\), \(\alpha_{\rm{M}O, \beta}\), and \(\alpha_{\rm IN,\beta}\), we apply the extended Sylvester's criterion, which states that a a Hermitian matrix is PSD if and only if all of its principal minors are nonnegative~\cite{gilbert1991positive}. 
Since \(\mathcal{\tilde{A}}\) is, by construction, block-diagonal, it is PSD if and only if both 
\(\mathcal{\til{A}}_{\rm{E}}\) and \(\mathcal{\til{A}}_{\rm{M}}\) are PSD. 
Accordingly, it suffices to apply the extended Sylvester's criterion independently to each \(3\times3\) dipolar block, 
requiring that all their principal minors are non-negative. 
Since \( \mathcal{\til{A}}_{\mathrm{E}} \) is a block-diagonal matrix incorporating an in-plane complex-rotation block, applying the extended Sylvester criterion straightforwardly yields the inequalities: \( \mathcal{J}_{\mathrm{E}} \ge |\mathcal{O}_{\mathrm{E}}| \ge 0 \) and \( \mathcal{T}_{\mathrm{E}} \ge 0 \). 

One could apply the extended Sylvester's criterion to \(\mathcal{\til{A}}_{\rm{M}}\) as we have just done for \(\mathcal{\til{A}}_{\rm{E}}\). 
However, \(\mathcal{\til{A}}_{\rm{M}}\) and \(\mathcal{\til{A}}_{\rm{E}}\) share the same algebraic form; hence, it can be shown that the conclusions obtained for \(\mathcal{\til{A}}_{\rm{E}}\) also apply to \(\mathcal{\til{A}}_{\rm{M}}\) under \(\rm{E} \rightarrow \rm{M}\).

Accordingly, we can summarize our findings on energy conservation for magneto-optical NPs as follows:
\begin{keyresultbox}
{Any magneto-optical NP satisfies  $\mathcal{J}_{\beta} \geq |\mathcal{O}_{\beta}| \geq 0$ and $\mathcal{T}_{\beta} \geq 0$ for both $\beta = \rm{E}, \rm{M}$. Thus, if $\mathcal{J}_{\beta} = 0$ then  $\mathcal{O}_{\beta} = 0$. The converse statement does not hold.  If the magneto-optical NP is lossless, then  $\mathcal{\tilde{A}}  = 0$, and thus,  Eqs.~\eqref{Abs_1}-\eqref{Abs_3} need to be simultaneously fulfilled for both $\beta = \rm{E}, \rm{M}$.}
\end{keyresultbox}

Importantly, our analytical results on energy conservation are not only crucial for physically consistent magneto-optical modelling, they also enable us to address a central open question in magneto-optics. This is: is does energy conservation preclude purely magneto-optical dipole scattering?
In other words, Can a magneto-optical NP exhibit $\alpha_{\rm{MO}, \beta} \neq 0$ with $\alpha_{\rm{OR}, \beta} = 0$? 
%In such a case, the scattered field produced by the NP would originate solely from the magneto-optical polarizability $\alpha_{\rm{MO}, \beta}$.
It should be highlighted that this magneto-optical response is not only conceptually striking but also actively sought after in the community as it would give rise, for instance, to effective magnetic monolopes~\cite{marques2024magneto}.
In this regard, it is also important to notice that the standard magneto-optically techniques, such as the Faraday rotation~\cite{faraday1857x, han2023strong}, Magneto-optical Kerr effects~\cite{kerr1877xliii, yang2022observation}, and Magnetic Circular Dichroism~\cite{stephens1970theory,eriksen2024chiral}, have been ultimately designed to detect the presence of $\alpha_{\rm{MO}, {\rm{E}}}$ and $\alpha_{\rm{MO}, {\rm{M}}}$ in the scattered electromagnetic field of magneto-optically active NPs.

However, our recent result given by  $\mathcal{J}_\beta \geq 0$ preclude the existence of purely magneto-optical responses conclusively: if $\alpha_{\rm{OR}, \beta} = 0$, then necessarily $\alpha_{\rm{MO}, \beta} = 0$ for both $\beta = \rm{E}, \rm{M}$ dipolar channels. This general finding follows directly from the non-negativity of the squared moduli of $\alpha_{\rm{MO}, \beta}$.

To date, however,  this finding has been overlooked in the literature due to the widespread practice of approximating the tensor \(\til{\alpha}_\beta\) given in Eq.~\eqref{pol_mag} to the first-order on the external magnetic field  \(\B_{\rm{ext}}\) ~\cite{, lodewijks2014magnetoplasmonic,pakdel2012faraday,maccaferri2013tuning,pineider2013circular,marinchio2014magneto,vincent2011magneto, edelstein2019magneto}.
In this (so-called)  linear regime in $\B_{\rm{ext}}$, $\alpha_{\rm{OR}, \beta}$ does not present magneto-optical terms ($\alpha_{\rm{OR}, \beta} \equiv \alpha_{\rm{IN}, \beta})$ and \(|\alpha_{\rm{MO}, \beta}|^2\) is omitted in Eq.~\eqref{H11}, namely, 
\begin{equation} \nonumber
\mathcal{J}_\beta \overbrace{\approx}^{\text{Linear regime}}
\Im \{ \alpha_{\rm{IN}, \beta} \} - \frac{k^3}{6 \pi}  |\alpha_{\rm{IN}, \beta}|^2.   
\end{equation}
This common approximation
makes Eq.~\eqref{H12} the only relation involving $\alpha_{\rm{MO}, \beta}$.
As a result, one might infer from Eq.~\eqref{H12} that a purely magneto-optical response is possible when $\alpha_{\rm{OR}, \beta} \equiv \alpha_{\rm{IN}, \beta} = 0$, provided \(\Re \{ \alpha_{\rm{MO}, \beta} \}  \neq 0\). 
Such an inference is incorrect and brings us to the following key result of this work: 
\begin{keyresultbox}
NPs cannot be purely magneto-optical (\(\alpha_{\rm{OR}, \beta} = 0\), \(\alpha_{\rm{MO}, \beta} \neq 0\)) without violating energy conservation. This general result holds for any material (lossless or lossy), electromagnetic size of the nanoparticle and strength and direction of the external magnetic field $\B_{\rm{ext}}$.
\end{keyresultbox}

The condition $\alpha_{\rm OR,\beta}=0$ forcing $\alpha_{\rm MO,\beta}=0$ mimics the behavior of chiral NPs~\cite{jaggard1979electromagnetic}.
In that work, it is shown that energy conservation also requires the chiral polarizability to vanish if either the electric or magnetic polarizability is zero~\cite{albooyeh2016purely}. Analogously, the Tellegen polarizability vanishes if the electric or magnetic polarizability is identically zero~\cite{tellegen1948gyrator}.
Taken together, we can state that energy conservation forbids purely bianisotropic (chiral and Tellegen) and magneto-optical responses at the single-particle level.

At this point, we  note that the previous energy-conservation constraint does not necessarily require \( |\alpha_{\rm{OR}, \beta}| > |\alpha_{\rm{MO}, \beta}| \), even though this inequality might be typically assumed in magneto-optics. We may then ask: does energy conservation, as expressed by Eq.~\eqref{def}, preclude \textit{strong} magneto-optics? In other words, is it possible to have \( |\alpha_{\rm{MO}, \beta}| \gg |\alpha_{\rm{OR}, \beta}| > 0 \)?

\emph{Strong Magneto-Optical Dipole Scattering.}--- Let us write the first Sylverster principal minor  $\mathcal{J}_{\beta} \geq 0$ as:
\begin{equation} \label{eq:S1_full}
 \Im \{ \alpha_{\rm{OR}, \beta} \} - \frac{k^3}{3 \pi} |\alpha_{\rm{OR}, \beta}|^2 \geq \frac{k^3}{6 \pi} \left(|\alpha_{\rm{MO}, \beta}|^2 - |\alpha_{\rm{OR}, \beta}|^2 \right).
\end{equation}
%The right-hand side of Eq.~\eqref{eq:S1_full} must be positive for $|\alpha_{\rm{MO}, \beta}|^2 > |\alpha_{\rm{OR}, \beta}|^2$ to hold. 
From Eq.~\eqref{eq:S1_full}, we immediately obtain the  threshold that $\alpha_{\rm{OR}, \beta}$ must satisfy in order to achieve $|\alpha_{\rm{MO}, \beta}|^2 > |\alpha_{\rm{OR}, \beta}|^2$:
\begin{equation} \label{eq:S1_threshold}
3 \pi \Im \{ \alpha_{\rm{OR}, \beta}  \} > k^3|\alpha_{\rm{OR}, \beta} |^2.
\end{equation}
Equation~\eqref{eq:S1_threshold} establishes a fundamental requirement for any single magneto-optical NP to reach $|\alpha_{\rm{MO}, \beta}| > |\alpha_{\rm{OR}, \beta}|$ for both $\beta = \rm{E}$, and  $\beta = \rm{M}$. We stress that this inequality is perfectly compatible with the conservation of energy.

In the following, we provide results for several magneto-optical NPs. For  simplicity, we consider sufficiently small NPs described by the 3x3 electric polarizability subset $\til{\alpha}_{\rm{E}}$.

\emph{Results.}--- A model of the dipolar magneto-optical response is needed for computing the polarizabilities $\alpha_{\rm{OR}, \rm{E}}$ and $\alpha_{\rm{MO}, \rm{E}}$. We use the Drude model for our next calculations~\cite{drude1900elektronentheorie, latella2025radiative, zhang2025observation, buddhiraju2020nonreciprocal, serrano2024collective}.
Assuming a linear response of the permittivity tensor $\til{\varepsilon}$ with  $\mathbf{B}_{\rm{ext}} = \pm B_0 \hat{\mathbf{e}}_z$, we have
\be \label{eps}
\til{\varepsilon} = 
\til{\varepsilon}(\B_{\text{ext}}) = \begin{pmatrix}
\varepsilon_{\rm{OR}} & \mp \im \varepsilon_{\rm{MO}}   &   0  \\ \pm \im \varepsilon_{\rm{MO}}    & \varepsilon_{\rm{OR}}&  0\\   0 &  0 &  \varepsilon_{\rm{IN}}
\end{pmatrix},
\ee 
where
\be \nonumber
\varepsilon_{\rm{OR}} &=&  \varepsilon_\infty 
+ \frac{\omega_p^2 (\omega + \im\gamma)}{\omega \left[\omega_c^2- (\omega + \rm{i} \gamma)^2 \right]}, 
\\ \nonumber
\varepsilon_{\rm{MO}} &=& - \frac{\omega_p^2 \omega_c}{\omega \left[\omega_c^2- (\omega + \rm{i} \gamma)^2 \right]},  \nonumber \qquad
\varepsilon_{\rm{IN}} = \varepsilon_\infty - \frac{\omega_p^2}{\omega(\omega + \rm{i}\gamma)}.
\ee
Here, \( \varepsilon_\infty \) is the high-frequency (background) permittivity; \( \omega_p  \) is the plasma frequency,  \( \omega_c  \) is the cyclotron frequency, and \( \gamma \) is the phenomenological damping rate.
Now, the polarizability tensor $\til{\alpha}_{\rm{E}}$ can be calculated from  Eq.~\eqref{eps}.
Following  Sáenz's notation~\cite{Albaladejo2010}, we write  $\til{\alpha}_{\rm{E}}$ {as} 
\be \label{pol}
\til{\alpha}_{\rm{E}} &=& \left( \til{\alpha}_0 ^{-1}-\frac{\im k^3}{6\pi}\til{\rm{I}}  \right)^{-1}, \\ \label{pol_0}
\til{\alpha}_0 &=& 3V\left(\til{\varepsilon}-\varepsilon_{\rm{h}}\til{\rm{I}}\right)\left(\til{\varepsilon}+2\varepsilon_{\rm{h}} \til{\rm{I}} -(x)^2 \left(\til{\varepsilon}-\varepsilon_{\rm{h}}\til{\rm{I}}\right)\right)^{-1}.
\ee
Here $k = n_{\rm{h}} \omega/c$ is the wavenumber of the incident electric field $\E_{\rm{inc}}$, $c$ is the speed of light,
$V=4\pi R^3/3$ is the volume of the spherical NP and  $x = kR$ its optical size. The mathematical derivation of $\til{\alpha}_{\rm{E}}$ given in Eq.~\eqref{pol}  is provided in Appendix~\ref{Appendix B}. 

In the following, we use Eq.~\eqref{pol} to compute the magneto-optical measures.
In particular, Fig.~\ref{fig} shows the response of a spherical magneto-optical NP with radius $R = 15~\mathrm{nm}$ and fixed parameters $\varepsilon_{\rm inf} = 1$ and $\hbar \omega_p = 4~\mathrm{eV}$. The top row displays the conservation of energy, quantified as $(\mathcal{J}_{\rm{E}} - |\mathcal{O}_{\rm{E}}|)/V$, for different damping rates $\hbar\gamma$. As shown in Figs.~\ref{fig}a–b, $(\mathcal{J}_{\rm{E}} - |\mathcal{O}_{\rm{E}}|)/V > 0$ for all considered values, confirming the generality of our energy-bounds relations for lossy magneto-optical NPs. Note that for $\hbar\gamma = 0$, Fig.~\ref{fig}c shows that $\mathcal{J}_{\rm{E}} = \mathcal{O}_{\rm{E}} = 0$, in perfect agreement with the lossless setting imposed by Eqs.~\eqref{Abs_1}–\eqref{Abs_3}.
Moreover, the bottom row of Fig.~\ref{fig}c presents $|\alpha_{\rm{MO}, \rm{E}}|^2 / |\alpha_{\rm{OR}, \rm{E}}|^2$ (log scale). As shown in these plots, $|\alpha_{\rm{MO}, \rm{E}}|^2 > |\alpha_{\rm{OR}, \rm{E}}|^2$ for different damping rates $\hbar \gamma$, as we have previously anticipated. As a matter of fact, one can obtain $|\alpha_{\rm{MO}, \rm{E}}|^2 / |\alpha_{\rm{OR}, \rm{E}}|^2 \sim 100$ for relatively low values of $\hbar \omega_c$; for instance, $\hbar \omega_c = 0.01$ eV provided that $\hbar \gamma \approx 0$.  

\begin{figure}
    \centering
\includegraphics[width=1\linewidth]{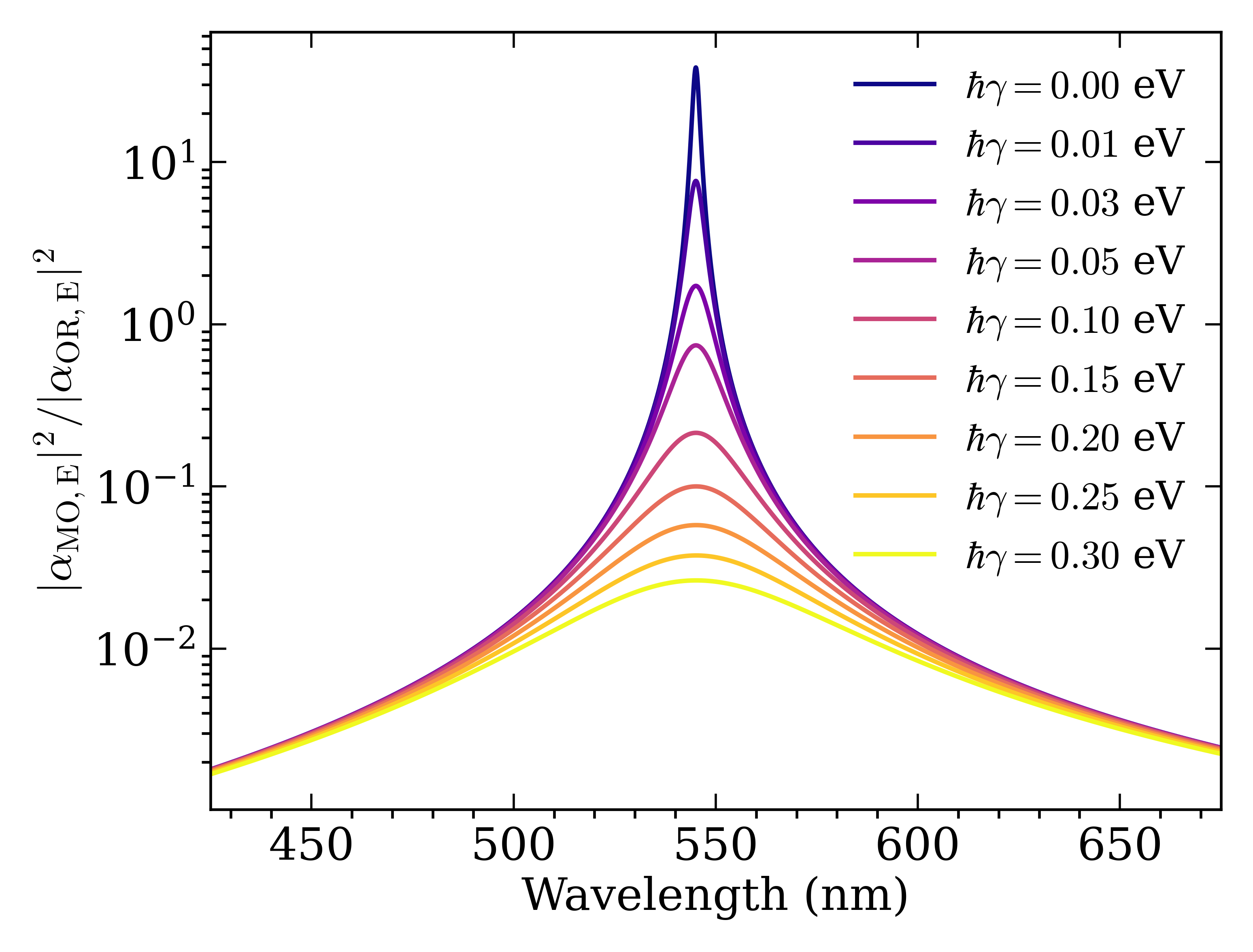}
    \caption{$|\alpha_{\rm{MO}, \rm{E}}|^2 / |\alpha_{\rm{OR}, \rm{E}}|^2$ (log scale) vs the wavelength for several constant values of the damping rates $\hbar \gamma$. The  parameters are fixed:  $\omega_c = 0.01$ eV, $R = 15~\mathrm{nm}$,  $\varepsilon_{\rm inf} = 1$ and $\hbar \omega_p = 4~\mathrm{eV}$.}
    \label{fig_3}
\end{figure}

To get further insights into this unexpected phenomenon, in Fig.~\ref{fig_3}, we depict $|\alpha_{\rm{MO}, \rm{E}}|^2 / |\alpha_{\rm{OR}, \rm{E}}|^2$ (log scale) vs several constant values of the damping rates $\hbar \gamma$ for the following fixed values: $\omega_c = 0.01$ eV, $R = 15~\mathrm{nm}$,  $\varepsilon_{\rm inf} = 1$ and $\hbar \omega_p = 4~\mathrm{eV}$. %As can be inferred for this plot, $|\alpha_{\rm{MO}, \rm{E}}|^2 / |\alpha_{\rm{OR}, \rm{E}}|^2$ reaches values much greater than , confirming that the magneto-optical polarizability can be significantly greater than the ordinary one.

In conclusion, we have derived the energy conservation law for axially-symmetric magneto-optical NPs described by electric and magnetic dipolar modes. This law is valid for both lossless and lossy materials, general incident illumination conditions, and external magnetic-field strengths, making it relevant for any scientific area dealing with the interaction of electromagnetic fields and magneto-optically active NPs. 

Importantly, our energy conservation expressions yield two key results:  pure magneto-optical dipolar scattering ($\alpha_{\rm OR}=0$, $\alpha_{\rm MO} \neq 0$) is precluded, while strong magneto-optical dipole scattering ($|\alpha_{\rm MO}| \gg |\alpha_{\rm OR}| > 0$) is not. 
These results not only establish fundamental limits in magneto-optics but also pave the way for realizing NPs exhibiting strong magneto-optical light scattering at the nanoscale. 

Finally, we should highlight that the constraints $\mathcal{J}_{\beta} \geq |\mathcal{O}_{\beta}| \geq 0$ and $\mathcal{T}_{\beta} \geq 0$ must be satisfied by the electric and magnetic dipolar coefficients of the magneto-optical Mie solution. Since no exact analytical expressions exist for these coefficients in the presence of an external magnetic field, these bounds impose strict conditions that any physically admissible solution must fulfill, thereby guiding the derivation of the exact magneto-optical Mie scattering coefficients.

\section*{Acknowledgements}
J.O-T. acknowledges fruitful discussions with Mr. Lukas Rebholz, Dr. Antonio García-Martín, and Dr. Manuel Marqués.
Moreover, J.O-T. acknowledges  financial support
from the Spanish Ministry of Science and Innovation
(MCIN) and AEI through Project
No. PID2022-137569NB-C43. J.O-T.  also acknowledges MICINN for the Ramon y Cajal Fellowship  (RYC2024-050342-I) and the fellowship (LCF/BQ/PR25/12110015) from “La Caixa” Foundation.

\clearpage

\section*{References}
\bibliography{Bib_tesis}
\clearpage

\onecolumngrid
\appendix
\section{Analytical derivation of $\til{\mathcal{A}}_\beta$} \label{Appendix A}
We present here the explicit analytical form of the Hermitian matrix $\til{\mathcal{A}}_\beta$ for an axially-symmetric magneto-optical particle, where $\beta = \rm{E}, \rm{M}$ denotes electric and magnetic dipolar responses, respectively. As in the main text, the $\til{\mathcal{A}}_\beta$ matrix is defined as
\begin{equation} \label{again}
\til{\mathcal{A}}_\beta = \frac{\til{\alpha}_\beta - \til{\alpha}_\beta^\dagger}{2 \im} - R_c \, \til{\alpha}_\beta^\dagger \til{\alpha}_\beta,
\end{equation}
with $R_c = k^3/(6 \pi)$, and the required polarizability tensors are given by
\begin{equation}
\til{\alpha}_\beta =
\begin{pmatrix}
\alpha_{\rm OR,\beta} & \mp \im \alpha_{\rm MO,\beta} & 0 \\
\pm \im \alpha_{\rm MO,\beta} & \alpha_{\rm OR,\beta} & 0 \\
0 & 0 & \alpha_{\rm{IN}, \beta}
\end{pmatrix}, \quad
\til{\alpha}_\beta^\dagger =
\begin{pmatrix}
\alpha_{\rm OR,\beta}^* & \mp \im \alpha_{\rm MO,\beta}^* & 0 \\
\pm \im \alpha_{\rm MO,\beta}^* & \alpha_{\rm OR,\beta}^* & 0 \\
0 & 0 & \alpha_{\rm{IN}, \beta}^*
\end{pmatrix}.
\end{equation}
As a reminded of the reader, here $\pm$ denotes the direction of application of the external magnetic field $\B_{\rm{ext}} = \pm B_0 \hat{\mathbf{e}}_{\rm{z}}$. Now, the first term of Eq.~\eqref{again} given by  $(\til{\alpha}_\beta - \til{\alpha}_\beta^\dagger)/(2 \im)$, can be expanded yielding
\begin{equation}
\frac{\til{\alpha}_\beta - \til{\alpha}_\beta^\dagger}{2 \im} =
\begin{pmatrix}
\Im(\alpha_{\rm OR,\beta}) & \mp \im  \Im(\alpha_{\rm MO,\beta}) & 0 \\
\pm \im  \Im(\alpha_{\rm MO,\beta}) & \Im(\alpha_{\rm OR,\beta}) & 0 \\
0 & 0 & \Im(\alpha_{\rm{IN}, \beta})
\end{pmatrix}.
\end{equation}
The term $\til{\alpha}_\beta^\dagger \til{\alpha}_\beta$ can be also computed explicitly. For the $2\times 2$ $xy$-block one finds
\begin{align}
(\til{\alpha}_\beta^\dagger \til{\alpha}_\beta)_{xx} &= |\alpha_{\rm OR,\beta}|^2 + |\alpha_{\rm MO,\beta}|^2, \\
(\til{\alpha}_\beta^\dagger \til{\alpha}_\beta)_{yy} &= |\alpha_{\rm OR,\beta}|^2 + |\alpha_{\rm MO,\beta}|^2, \\
(\til{\alpha}_\beta^\dagger \til{\alpha}_\beta)_{xy} &= \mp 2 \im  \Re(\alpha_{\rm OR,\beta}^* \alpha_{\rm MO,\beta}), \\
(\til{\alpha}_\beta^\dagger \til{\alpha}_\beta)_{yx} &= \pm 2 \im \, \Re(\alpha_{\rm OR,\beta}^* \alpha_{\rm MO,\beta}).
\end{align}
The uncoupled $zz$ component reads $(\til{\alpha}_\beta^\dagger \til{\alpha}_\beta)_{zz} = |\alpha_{\rm{IN}, \beta}|^2$, and the remaining off-diagonal terms vanish. Now, combining both contributions yields the Hermitian matrix
\begin{equation}
\til{\mathcal{A}}_\beta =
\begin{pmatrix}
\Im(\alpha_{\rm OR,\beta}) - R_c(|\alpha_{\rm OR,\beta}|^2 + |\alpha_{\rm MO,\beta}|^2) & \mp \im \left(\Im(\alpha_{\rm MO,\beta}) - 2 R_c \Re(\alpha_{\rm OR,\beta}^* \alpha_{\rm MO,\beta})\right) & 0 \\
\pm \im \left(\Im(\alpha_{\rm MO,\beta}) - 2 R_c \Re(\alpha_{\rm OR,\beta}^* \alpha_{\rm MO,\beta})\right) & \Im(\alpha_{\rm OR,\beta}) - R_c(|\alpha_{\rm OR,\beta}|^2 + |\alpha_{\rm MO,\beta}|^2) & 0 \\
0 & 0 & \Im(\alpha_{\rm{IN}, \beta}) - R_c |\alpha_{\rm{IN}, \beta}|^2
\end{pmatrix},
\end{equation}
As in the main text,  we can define the quantities: 
\be \label{H11_a}
\mathcal{J}_{\beta} &=&  
\Im(\alpha_{\rm{OR},\beta})
- \frac{k^3}{6\pi} \left( |\alpha_{\rm{OR},\beta}|^2 + |\alpha_{\rm {MO}, \beta}|^2 \right),
\\  \label{H12_a}
\mathcal{O}_{\beta} &=& 
\Im(\alpha_{\rm {MO}, \beta})
- \frac{k^3}{3\pi} \Re(\alpha^*_{\rm{OR},\beta} \alpha_{\rm {MO}, \beta}),
\\  \label{H33_a}
\mathcal{T}_{\beta} &=& 
\Im(\alpha_{\rm{IN}, \beta})
- \frac{k^3}{6\pi} |\alpha_{\rm{IN}, \beta}|^2.
\ee
Accordingly, we can express $\til{\mathcal{A}}_\beta$ as 
\begin{equation} \label{def_app}
\mathcal{\til{A}}_\beta =
\begin{pmatrix}
\mathcal{J}_{\beta} & \mp \im \mathcal{O}_{\beta} & 0 \\
\pm \im \mathcal{O}_{\beta} & \mathcal{J}_{\beta} & 0 \\
0 & 0 & \mathcal{T}_{\beta}
\end{pmatrix}.
\end{equation}
This last expression is identical to Eq.~\eqref{def} of the main text.

\section{Derivation of the magneto–optical polarizability tensor} \label{Appendix B}
In this Appendix, we derive the polarizabilities of a magneto-optical spherical NP in the Rayleigh approximation, including the radiative correction and the depolarization factor. Our starting point is
Eq.~\eqref{eps} of the main text, which reads
\begin{equation} \label{epsilon_reduced}
\til{\varepsilon} =
\begin{pmatrix}
\varepsilon_{\rm{OR}} & \mp \mathrm{i}\varepsilon_{\mathrm{MO}} & 0 \\
\pm \mathrm{i}\varepsilon_{\mathrm{MO}} & \varepsilon_{\rm{OR}} & 0
\\ 0 & 0 & \varepsilon_{\rm{IN}}
\end{pmatrix}.
\end{equation}
Inserting Eq.~\eqref{epsilon_reduced} into Eq.~\eqref{pol_0} of the main text yields the following 2x2 tensor
\begin{equation}
\til{\alpha}_{0,xy} =
\begin{pmatrix}
\alpha_{0,xx} & \alpha_{0,xy}\\[2pt]
-\alpha_{0,xy} & \alpha_{0,xx}
\end{pmatrix},
\end{equation}
where 
\begin{align}
\label{eq:alphaxx_x2}
\alpha_{0,xx} &=
\frac{
-3 V (\varepsilon_{\rm{h}} - \varepsilon_{\rm OR}) \left[(2 + x^2) \varepsilon_{\rm{h}} + (1 - x^2) \varepsilon_{\rm OR}\right] - 3 V (1 - x^2) \varepsilon_{\rm MO}^2
}{
\left[(2 + x^2) \varepsilon_{\rm{h}} + \varepsilon_{\rm OR} (1 - x^2) \right]^2 - (x^2 - 1)^2 \varepsilon_{\rm MO}^2
},
\\
\label{eq:alphaxy_x2}
\alpha_{0,xy} &= 
 \frac{\mp 9\mathrm{i}V \varepsilon_{\rm{h}}\varepsilon_{\mathrm{MO}}}{\left[(2+x^{2})\varepsilon_{\rm{h}}+\varepsilon_{\rm OR} (1 - x^2)  \right]^2 -\left(x^2-1 \right)^2 \varepsilon_{\mathrm{MO}}^{2}}.
\end{align}
The decoupled $z$ component is given by 
\begin{equation}
\label{eq:alphazz_x2}
\alpha_{0,zz}=3V\frac{\varepsilon_{\rm{IN}}-\varepsilon_{\rm{h}}}{\varepsilon_{\rm{IN}}+2\varepsilon_{\rm{h}} - x^{2}(\varepsilon_{\rm{IN}}-\varepsilon_{\rm{h}})}.
\end{equation}
The radiation--corrected (dynamic) polarizability is
\begin{equation}
\label{eq:alphaE_def_x2}
\til{\alpha}_{\mathrm{E}}=\big(\til{\alpha}_0^{-1}- R_{\rm{ic}}\til{\rm{I}}\big)^{-1},
\qquad
R_{\rm{ic}}=\frac{\mathrm{i}k^{3}}{6\pi} =  \im R_c,
\end{equation}
where $k$ is the wavenumber in the host medium.
For the $xy$ block, the radiation-corrected components are
\begin{align}
\label{eq:alphaxx_dyn_x2}
\alpha_{xx} &= \alpha_{yy}
= \frac{\alpha_{0,xx} - R_{\rm{ic}}(\alpha_{0,xx}^{2}+\alpha_{0,xy}^{2})}{1 - 2R_{\rm{ic}} \alpha_{0,xx} + R_{\rm{ic}}^{2}(\alpha_{0,xx}^{2}+\alpha_{0,xy}^{2})},
\\
\label{eq:alphaxy_dyn_x2}
\alpha_{xy} &= -\alpha_{yx}
= \frac{\alpha_{0,xy}}{1 - 2R_{\rm{ic}} \alpha_{0,xx} + R_{\rm{ic}}^{2}(\alpha_{0,xx}^{2}+\alpha_{0,xy}^{2})}.
\end{align}
Now, the $zz$ component reads as
\begin{equation}
\label{eq:alphazz_dyn_x2}
\alpha_{zz}=\frac{\alpha_{0,zz}}{1-R_{\rm{ic}}\alpha_{0,zz}}.
\end{equation}
Hence, the 3x3 electric polarizability tensor is
\begin{equation}
\label{eq:alphaE_tensor_x2}
\til{\alpha}_{\mathrm{E}} =
\begin{pmatrix}
\alpha_{xx} & \alpha_{xy} & 0\\[2pt]
-\alpha_{xy} & \alpha_{xx} & 0\\[2pt]
0 & 0 & \alpha_{zz}
\end{pmatrix},
\end{equation}
with $\alpha_{0,xx}$ and $\alpha_{0,xy}$ defined by Eqs.~\eqref{eq:alphaxx_x2}--\eqref{eq:alphaxy_x2} and $\alpha_{0,zz}$ is given in Eq.~\eqref{eq:alphazz_x2}. By inspecting Eq.~\eqref{pol_mag} of the main text, we realize that  $\alpha_{\rm{xx}} = \alpha_{\rm{OR}, \rm{E}}$ and  $\alpha_{xy} = \mp \im \alpha_{\rm{MO} ,\rm{E}}$ and $\alpha_{\rm{zz}} = \alpha_{\rm{IN}}$. Note that this leaves $\alpha_{\rm{MO}, \rm{E}}$ independent of the direction of the external magnetic field, as required to compute the energy-conservation bounds given in the main text.

\end{document}